\documentclass[11pt,twoside]{article}
\usepackage{asp2004}
\usepackage{epsf}
\usepackage{graphics}
\usepackage{amsmath,epsfig,rotating,amssymb}
\usepackage{lscape}
\def\edcomment#1{\iffalse\marginpar{\raggedright\sl#1\/}\else\relax\fi}
\reversemarginpar

\begin{document}
\title{On the structure of the interstellar atomic gas}
 \author{P. Hennebelle$^1$ and E. Audit$^2$}
\affil{$^1$ Laboratoire de radioastronomie millim{\'e}trique, UMR 8112 du CNRS, 
\newline {\'E}cole normale sup{\'e}rieure et Observatoire de Paris, 24 rue Lhomond,
\newline 75231 Paris cedex 05, France  
\newline $^2$ Service d'Astrophysique, CEA/DSM/DAPNIA/SAp, C. E. Saclay,
\newline F-91191 Gif-sur-Yvette Cedex}

\begin{abstract}
The interstellar atomic hydrogen is known to be a 2-phase medium in which turbulence 
plays an important r\^ole. Here we present high resolution numerical simulations
describing the gas from tens of parsec down to hundreds of AU. This high resolution
allows to probe numerically, the small scale structures which naturally arises 
from the turbulence and the 2-phase physics. 
\end{abstract}

\vspace{-0.5cm}
\section{Introduction}
The interstellar atomic hydrogen  (HI)   has been extensively observed over the years 
(e.g.  Kulkarni \& Heiles 1987, Dickey \& Lockmann 1990, 
Joncas et al. 1992, Heiles \& Troland 2003, 2005, Miville-Desch\^enes et al. 2003). 
Although HI has also received a lot of attention from the theoretical point of view
(Field 1965, Field et al. 1969, Wolfire et al. 1995)
and in spite of the  early recognition that HI is a turbulent medium
(e.g. Heiles \& Troland 2005),  it is only recently that the dynamical 
properties of HI have been  investigated mainly because of the large dynamical range 
which is necessary to treat the problem. 

Here we present high resolution numerical simulations aiming to describe a turbulent 
atomic hydrogen flow from scales of few tens of parsec down to scales of few hundreds of AU.
Such simulations have been performed with less spatial resolution by Gazol et al. (2001), 
Audit \& Hennebelle (2005), Heitsch et al. (2005, 2006), V\'azquez-Semadeni et al. (2006). 
In section 2, we first present the equations of the problem, we then discuss the 
different spatial scales important in this problem as 
well as the numerical setup. In section 3, we present our results. Section 4 concludes 
the paper. 

\begin{figure}[!ht]
\includegraphics[width=15cm,angle=90]{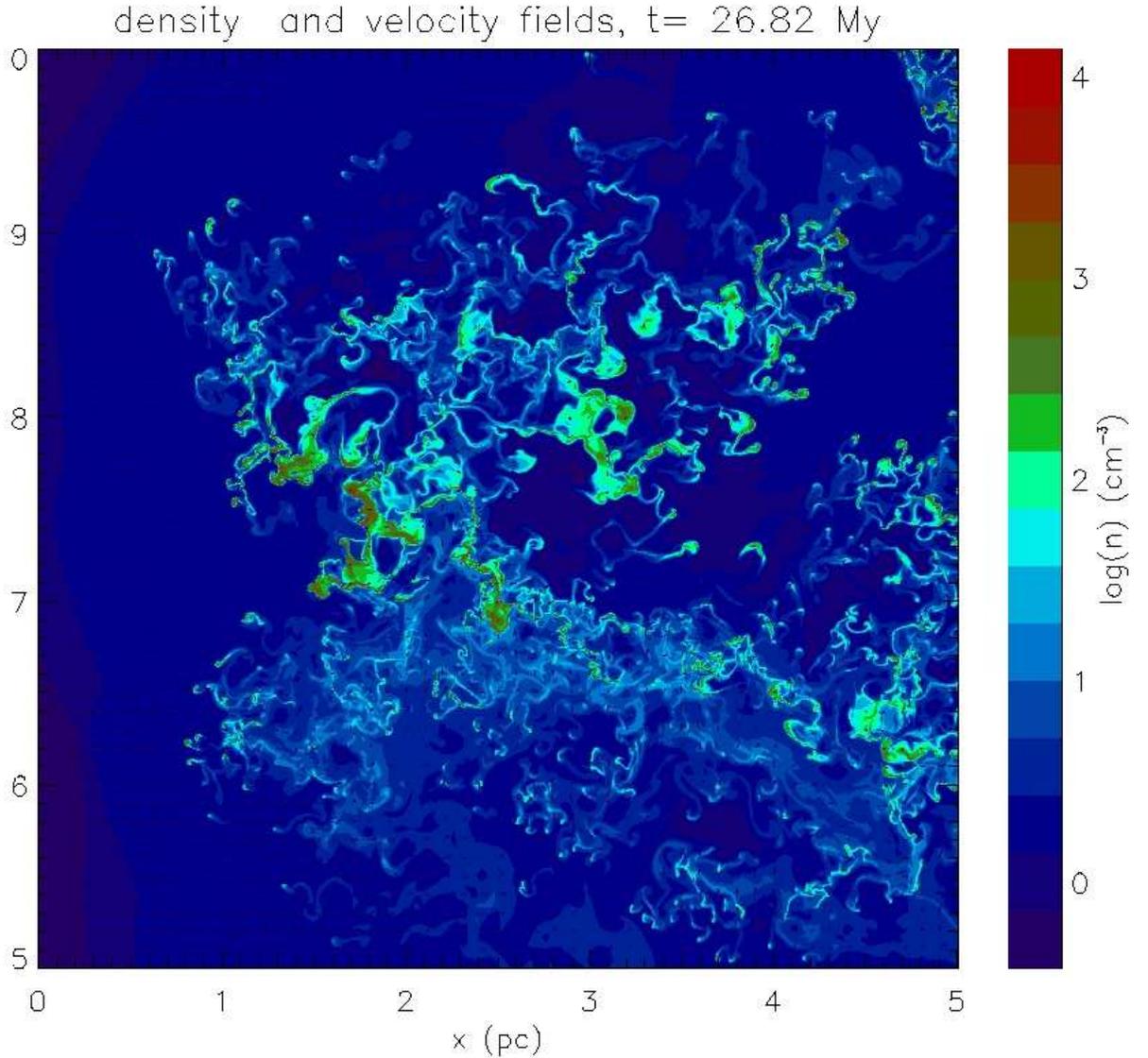}
\caption{Density and velocity fields at time 26.82 Myrs. }
\label{bigchamps}
\end{figure}

\begin{figure}[!ht]
\includegraphics[width=15cm,angle=90]{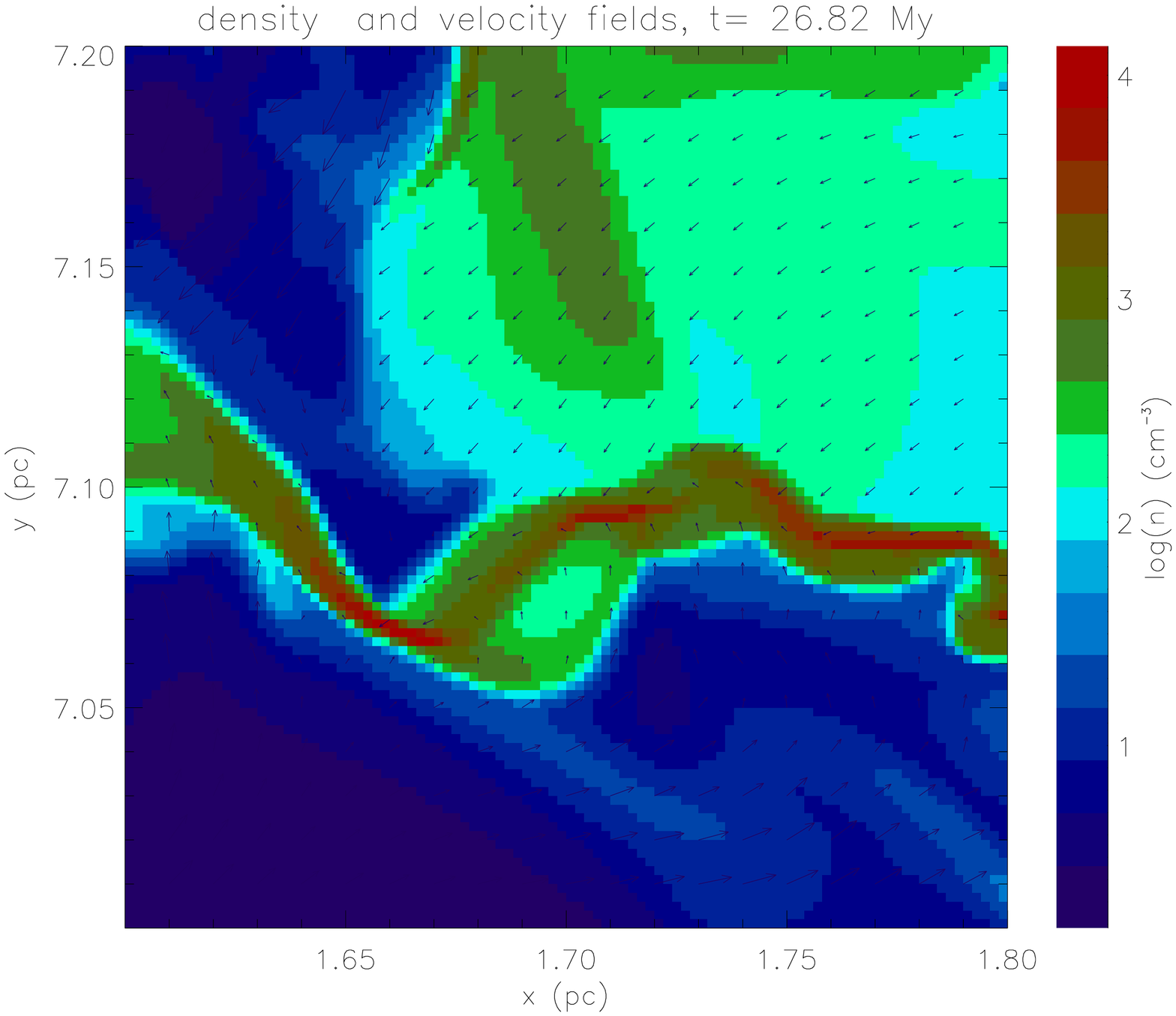}
\caption{Spatial zoom of Fig.~\ref{bigchamps}. }
\label{smallchamps}
\end{figure}

\begin{figure}[!ht]
\includegraphics[width=10cm]{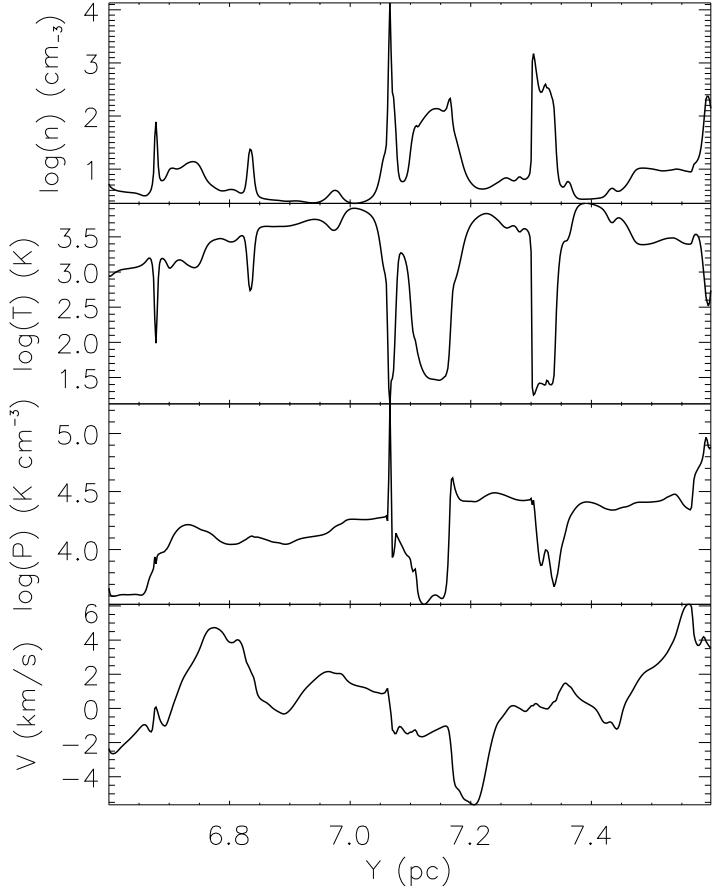}
\caption{One dimensional cut along y-axis of the field displayed in Fig.~\ref{bigchamps}.
Density, temperature, Pressure and velocity fields are displayed.}
\label{coupe}
\end{figure}


\section{Equations, Spatial scales and numerical Setup}

We consider the usual fluid equations for a radiatively cooling gas  
including thermal conductivity namely, 
\begin{alignat}{4}
\label{mcons}
\partial_t \rho       & \ +   \nabla . [\rho u]              & =  & 0, \\
\label{momcons}
\partial_t \rho u     & \ +   \nabla . [\rho u\otimes u + P] & =  & 0, \\
\label{econs}
\partial_t  E         & \ +   \nabla . [u(E + P)]            & =  & - {\cal L}(\rho,T) + \nabla ( \kappa(T) \nabla T ).
\end{alignat}
$\rho$ is the mass density, $u$ the  velocity, $P$ the pressure,
$E$ the  total energy and  ${\cal L}$  the  cooling function which 
includes Lyman-$\alpha$  $C^+$ and $O$ line cooling and grains photoelectric
heating.   
The gas is  assumed to be a perfect  gas with $\gamma = 5/3$
and with a mean molecular weight $\mu = 1.4 m _H$, where $m _H$ is
the mass of the proton.
$\kappa$ is the thermal conductivity and is given by 
$\kappa(T) = \gamma C_v \eta(T)$ where $C_v=k_b / m_H / (\gamma -1)$,
 $\eta=5.7 \times 10 ^ {-5}$ (T/1 K)$^ {1/2}$ g cm$^{-1}$ s$^{-1}$ and $k_b$ is the 
Boltzman constant. Various physical lengths appear to play an important r\^ole for the 
structure and the dynamics of this 2-phase flow.

The Field length (Field 1965, Koyama \& Inutsuka 2004) 
is the length at which thermal 
diffusivity becomes comparable to the heating and cooling 
term. Its typical value is about 0.1 pc in the WNM and 
about $10 ^{-3}$ pc in the CNM. The Field length is also 
the typical scale of the thermal fronts which connect
the cold and warm phases.

Three scales whose origin is due to dynamical 
processes have to be distinguished. 
The first scale is the cooling length of the WNM, i.e. the product 
of the cooling time and the sound speed within WNM, 
$\lambda_{\rm cool}= \tau_{\rm cool} \times C_{\rm s,WNM}$. This scale is about 10 pc
and corresponds to the typical length at which WNM is non linearly unstable
and can be dynamically triggered by a compression into the unstable regime
(Hennebelle \& P\'erault 1999, 2000, Koyama \& Inutsuka 2000).

Since the ratio between the CNM and WNM  density is about hundred, the
size of a CNM  structure formed through the  contraction of a piece of
WNM of  size  $\lambda_{\rm  cool}$,  will be  typically hundred  time
smaller (assuming a monodimensionnal  compression). Therefore the size
of the CNM structures is  about the cooling  length of WNM divided  by
hundred, leading to about $\lambda_{\rm cool}/100 \simeq $0.1 pc.

As first pointed out by Koyama \& Inutsuka (2002), the fragments
of CNM have a velocity dispersion with respect to each others which is a 
fraction of the sound speed of the medium in which they are embedded, i.e. 
WNM. Since the contrast between the sound speed within the two phases is about 
10, it means that CNM structures undergo collisions at Mach number, 
$M \simeq$  10. 
If for simplicity we assume isothermality and apply simple 
Rankine-Hugoniot conditions, we obtain that the size of the shocked 
CNM structures is given by the size of the CNM structures divided by 
$M^2$ which is about $10^{-3}$ pc.

As clearly shown by these numbers, the resolution necessary to describe 
fairly a 2-phase fluid like the neutral atomic gas is very mandatory. 
We therefore choose to perform 2D numerical simulations which allow to 
describe much smaller scales than 3D simulations. We use 10000$^4$ 
cells with a 20 pc size box leading to a spatial resolution of 0.002 pc.

The boundary conditions consist in an imposed converging flow at the left and
right faces of amplitude $1.5 \times C_{s,wnm}$ on top of which turbulent
fluctuations have been superimposed.
On top and bottom face outflow conditions have been setup. This means that 
the flow is free to escape the computational box across these 2 faces.
The initial conditions are a uniform low density gas 
($n \simeq 0.8$ cm$^{-3}$) at thermal  equilibrium corresponding to WNM. 
It is worth noticing that initially no CNM is present in the computational 
box. The simulations are then runned until a statistically stationary state 
is reached. This requires typically about 5 to 10 box crossing times.
At this stage the incoming flow of matter compensate on average the 
outgoing material.

\section{Results}
Figure~\ref{bigchamps} shows the density field at time 26.82 Myrs after the beginning
of the calculation. At this stage the statistical properties of the flow do not evolve 
any more. The structure of the medium appears to be rather complex. The cold phase is 
very fragmented into several long living cloudlets confined by the external pressure of 
the WNM. This is more clearly seen in Fig.~\ref{smallchamps} and~\ref{coupe} in which  
much smaller scales are visible. 
The importance of  the pressure confinement of the  structures by the 
surrounding warm gas is well seen on the density peaks of Fig.~\ref{coupe} which, 
except for the 
notable exception of the structure located at $x \simeq 7.05$ pc, show either no correlation
with the pressure or an anti-correlation. Large density fluctuations, well above the 
mean CNM density, can be seen. Such fluctuations are due to the collisions between the cold CNM 
fragments and are therefore located at the stagnation points of converging 
flows.
 This is well illustrated by the small (400-800 AU) and dense structures seen 
in Fig.~\ref{smallchamps} at $x \simeq 1.66, y \simeq 7.07$ pc, 
$x \simeq 1.72, y \simeq 7.10$ pc and $x \simeq 1.77, y \simeq 7.90$ pc for 
which the density reaches about 10$^4$ cm$^{-3}$.
  The size and density of the structures is therefore comparable 
  to the typical  values of shocked CNM 
fragments recalled in section 2 and are reminiscent of the values inferred for the 
tiny small scale atomic structures (TSAS) quoted e.g. by Heiles (1997). It is therefore 
tempting to propose that the TSAS are naturally produced by the 2-phase nature of the flow
which naturally generates high Mach collisions between cold CNM fragments.
Note that the spatial resolution available in the simulation presented here, is still not 
sufficient to fully resolve these small structures and one expects that with higher 
resolution, structures with even larger densities and smaller size will be formed.   
As can be seen in Fig.~\ref{coupe} ($y \simeq 7.05$), the pressure in these objects is 
well correlated with the density as it is expected in a high Mach number shock. This large 
pressure is dynamically maintained by the ram pressure of the large scale converging flow and the life time of the shocked CNM structure is about 
the size of the colliding CNM clouds divided by their respective velocity. For 0.1 pc  clouds endorging collisions at Mach 10, this is about 10$^4$ years.

Another interesting aspect is the large number of small  CNM structures seen in 
Fig.~\ref{bigchamps} and~\ref{coupe} (e.g. $x\simeq 6.67$ and $x \simeq 6.83$ pc). 
These structures which have a density of about 100 cm$^{-3}$ can be as small 
as few thousands of AU. They are reminiscent of recent observations of low column 
density CNM structures observed by Braun \& Kanekar (2005) and 
Stanimirovi\'c \& Heiles (2005). 

Quantitative statistical characterizations of these numerical simulations,
 as well as more quantitative comparisons with observations, 
will be given in Hennebelle \& Audit (2007) and 
Hennebelle, Audit \& Miville-Desch\^enes (2007).

Finally we note that in order to verify that the trends inferred from these simulations 
is not affected qualitatively by their bidimensionality, we have performed 3D simulations 
with a grid of 1200$^3$ cells. Although the resolution of these simulations 
is not sufficient to provide a 
description of the CNM structures down to the smaller scales, we have obtained 
similar results.

\section{Conclusion}
We have performed high resolution numerical simulations of a turbulent atomic gas. 
The resulting structure appears to be very complex. The CNM is very fragmented 
in long living cloudlets bounded by contact discontinuities and confined by the 
surrounding WNM. 
Small scale structures either dense (up to 10$^4$ cm$^{-3}$) or having density 
of standard CNM, appear to form naturally as a consequence of the turbulence and the 
2-phase physics. Both have physical parameters which are reminiscent with the values
inferred from observations of TSAS and recently observed low column density atomic clouds.

\end{document}